\documentclass[12pt]{article}
\usepackage{cite}
\usepackage[T1]{fontenc} % Enable T1 encoding for proper font support
\usepackage[utf8]{inputenc} % Ensure UTF-8 input encoding
\usepackage{amsmath,amssymb,amsfonts}
\usepackage{algorithm}
\usepackage{algpseudocode}
\usepackage{algorithmicx}
\usepackage{caption}
\usepackage{subcaption}
\usepackage{graphicx}
\usepackage{braket}
\usepackage{multirow}
\usepackage[normalem]{ulem}
\usepackage{setspace}
\usepackage[a4paper,margin=1in]{geometry}
\usepackage[
    colorlinks=true,
    pdfborder={0 0 0},
    allcolors=blue
]{hyperref}

\title{Enhancing Knapsack-based Financial Portfolio Optimization Using Quantum Approximate Optimization Algorithm}

\author{
    Chansreynich Huot\textsuperscript{1}, Kimleang Kea\textsuperscript{1}, \\
    Tae-Kyung Kim\textsuperscript{2}, and Youngsun Han\textsuperscript{1*}
}

\date{}

\begin{document}

\maketitle
\begin{center}
\textsuperscript{1}Department of AI Convergence, Pukyong National University, Nam-gu, Busan 48513, South Korea \\
\textsuperscript{2}Department of Management Information Systems, Chungbuk National University,  1, Chungdae-ro, Seowon-gu, Cheongju-si, Chungcheongbuk-do, South Korea\\
\vspace{6pt}
\noindent *\textbf{Corresponding author:} Youngsun Han (\texttt{youngsun@pknu.ac.kr})
\end{center}

\vspace{10pt}

\begin{abstract}
Portfolio optimization is a primary component of the decision-making process in finance, aiming to tactfully allocate assets to achieve optimal returns while considering various constraints. Herein, we proposed a method that uses the knapsack-based portfolio optimization problem and incorporates the quantum computing capabilities of the quantum walk mixer with the quantum approximate optimization algorithm (QAOA) to address the challenges presented by the NP-hard problem. Additionally, we present the sequential procedure of our suggested approach and demonstrate empirical proof to illustrate the effectiveness of the proposed method in finding the optimal asset allocations across various constraints and asset choices. Moreover, we discuss the effectiveness of the QAOA components in relation to our proposed method. Consequently, our study successfully achieves the approximate ratio of the portfolio optimization technique using a circuit layer of $p \geqslant 3$, compared to the classical best-known solution of the knapsack problem. Our proposed methods potentially contribute to the growing field of quantum finance by offering insights into the potential benefits of employing quantum algorithms for complex optimization tasks in financial portfolio management.
\end{abstract}

\section*{Keywords}
best-known solution (BKS), portfolio optimization, knapsack problem, quantum approximate optimization algorithm (QAOA)

\maketitle

\section{Introduction}

\label{sec:introduction}
The financial industry plays a remarkable role in the economic health and growth of every country. In the constantly evolving world of financial markets, technological advancements are shaping traditional portfolio optimization methods, presenting new opportunities and challenges; this process involves enlarging assets to reduce risk by offsetting individual risk profiles crucial to the investment process~\cite{speidell1989po}. Although there are many portfolio optimization models, they possibly have limitations. For example, the mean-variance model, which provides solutions as percentages of the total budget, can result in fractional allocations of non-feasible assets~\cite{vaezi2019portfolio}. In the constant pursuit of optimization, some studies have explored extensions to portfolio optimization~\cite{loke2023portfolio}. Among these, a technique that reformulates the financial portfolio optimization problem as a knapsack-based problem has been proposed~\cite{martello1987algorithms, mansini2015la}. 

Knapsack problem~\cite{vaezi2019portfolio} was adapted for portfolio optimization by treating all assets included in the portfolio as items. The value of each asset is represented by its expected return, usually estimated using historical data or forecasting techniques. The weight of each item corresponds to a risky asset, typically measured by its standard deviation or the variance-covariance matrix of the assets. The knapsack capacity corresponds to the budget or available capital that can be invested in the portfolio. By reformulating the knapsack problem, the primary objective is to maximize the total value of the portfolio while satisfying the capacity constraint and potentially other constraints, such as a target return or a minimum number of assets. However, similar to~\cite{chu1998genetic}, the knapsack problem is considered NP-hard even with polynomial-bounded weights and values. Therefore, exploring new paradigms for optimization owing to the computational complexity of solving this problem is crucial~\cite{honggang2009quantum, gunjan2024quantum}.

Similarly, the paradigm of quantum computing, leveraging the properties of quantum mechanics, has been developed to solve complex problems intractable for classical computers~\cite{herman2023quantum, cao2024softwarized}. The financial domain is a primary aspect of quantum computing, with applications spanning price derivation, risk modeling, portfolio optimization, and fraud detection~\cite{saxena2023financial}. As technology redefines problem-solving in finance, the quantum approximate optimization algorithm (QAOA), rooted in quantum computing, offers the promise of efficiently finding approximate solutions for computationally demanding problems within the polynomial-bounded NP optimization complexity class~\cite{farhi2014quantum}. In~\cite{van2021quantum,kea2023leveraging}, the application of QAOA to the knapsack problem is explored for optimization purposes.

In this paper, we reformulate the classical portfolio optimization problem as a knapsack problem that allows for an efficient mapping onto quantum algorithms, particularly advantageous for NP-hard financial optimization. By treating each asset's expected return as value and its risk as weight, the knapsack model aligns portfolio constraints with total risk, permitting optimization under capacity constraints. Our approach introduces a novel Quantum Walk Mixer (QWM) combined with the Quantum Approximate Optimization Algorithm (QAOA), integrating the Quantum Fourier Transform (QFT) within the oracle. This QWM-QAOA framework ensures feasible states are targeted while reducing computational complexity. Empirical results demonstrate that QWM-QAOA enhances solution space exploration, providing robust portfolio configurations with high approximation ratios even under various noise conditions, thus showcasing a significant advancement in leveraging quantum algorithms for practical finance applications. The contributions of our study can be summarized as follows:
\begin{itemize}
    \item We propose a portfolio selection approach that formulates the portfolio problem as a knapsack concern by incorporating expected returns from the Markowitz model~\cite{markowitz1976markowitz} and setting the capacity according to the knapsack framework.
    \item We present a Quantum Walk Mixer to Quantum Approximate Optimization Algorithm (QWM-QAOA) for the knapsack problem, integrating a shallow circuit layer to decrease computational complexity and improve solution quality.
    \item We demonstrate that our model consistently enhances the identification of optimal solutions for the knapsack problem, achieving an impressive approximation ratio ranging from 100\% to 95\% in scenarios involving the selection of 2 to 8 stocks.
\end{itemize}

The paper is organized as follows: Section~\ref{sec:background} provides the background knowledge. Section~\ref{sec:related} explores the related work. Section~\ref{sec:proposed} details our proposed method. Section~\ref{sec:evaluation} demonstrates the experimental setup and performance evaluation. Section~\ref{sec:discussion} dissects the findings of the proposed method, underscoring its achievements and limitations. Finally, we conclude our study in Section~\ref{sec:conclusion}.

\section{Background}
\label{sec:background}
In this section, we briefly describe the background knowledge of portfolio optimization, the knapsack problem, and the quantum approximate optimization algorithm (QAOA).

\subsection{Portfolio Optimization}
Portfolio optimization is a mathematical framework that maximizes returns while minimizing risks through strategically selecting assets within an investment portfolio~\cite{anagnostopoulos2010portfolio}. This is typically achieved by strategically allocating the proportion of each asset in the portfolio to optimize the risk-return tradeoff by considering the specified risk tolerance. The process involves four steps: i) identifying suitable assets, ii) projecting anticipated yields based on historical data for future forecasts, iii) quantifying the risk by assessing the uncertainty of each asset, and iv) selecting the optimal portfolio that maximizes the expected yield for a given risk level~\cite{gunjan2023brief}. One of the various models used in this study was the Markowitz model developed by Harry Markowitz in 1952~\cite{markowitz1976markowitz}. This analysis is in conjunction with the variance of the rate of return, providing a significant assessment of portfolio risk under a rational framework of assumptions. The Markowitz model represented the maximum expected yield by allocating funds into stocks as follows~\cite{kamil2006portfolio}:
\begin{equation}
    R_{i}  = \sum _{t=1}^{\infty } d_{i}{}_{t} r_{i}{}_{t} ,
\end{equation}
where ${r_i}_t$ indicates the anticipated return at time $t$ per stock invested in, and ${d_i}_t$ is the rate at which return in the $i_{th}$ security where time $t$ is discounted back to the presents. The standard deviation—the variance of return—is a statistical measure used as an indicator of the uncertainty or risk linked to return. These statistical indicators effectively measure the extent to which returns deviate unpredictably from the average value over a specific period. The variance represents the degree of variation exhibited by the return $R_i$ concerning the expected return $[E(R_i)]$, as illustrated:  
\begin{equation}
    \sigma _{i}^{2}  = \frac{1}{N}\sum _{i=1}^{N}[ R_i - E( R_{i})]^{2} .
\end{equation}
The covariance of returns measures the relative riskiness of a security within a portfolio of securities. For two securities, denoted as $i$ and $j$, the covariance of their returns as 
\begin{equation}
    \sigma _{i}{}_{j}  = E\{[ R_{i}  - E( R_{i})][ R_{j} -E( R)]\}  .
\end{equation}
Furthermore, covariance can be measured depending on the variability of the two individual return series:
\begin{equation}
     \rho _{i}{}_{j}  = \frac{\sigma _{i}{}_{j}}{\sigma _{i} \sigma _{j}} ,
\end{equation}
where ${\rho_i}_j$ is the correlation coefficient of returns, and $\sigma_i$ and $\sigma_j$ are the standard deviation of $R_{it}$ and $R_{jt}$. As previously stated, an efficient portfolio is characterized by selecting individual assets within the portfolio and weighting each asset. Therefore, the portfolio return is calculated as a weighted average of the returns of the individual investments within the portfolio. Next, $X_i$ denoted as the weight and applied to each return of the portfolio takes form as follows: 
\begin{equation}
     \begin{array}{l}
R_{p}  = \sum _{t=1}^{\infty }\sum _{i=0}^{N} d_{i}{}_{t} r_{i}{}_{t} X_{i}\\
\\
= \sum _{i=0}^{N} X_{i}\left(\sum _{t=1}^{\infty } d_{i}{}_{t} r_{i}{}_{t}\right)\\
\\
=  \sum _{i=0}^{N} X_{i} R_{i} ,
\end{array}
\end{equation}
where $R_i$ is independent of $X_i$. The simplified version of the variance of a portfolio can be written as  
\begin{equation}
\sigma _{p}^{2}  = \sum _{i=0}^{N} X_{i}^{2} \sigma _{i}^{2}   + \sum _{i=0}^{N}\sum _{j=1}^{N} X_{i} X_{j} \sigma _{i}{}_{j},   
\end{equation}
where $\sigma_{p}^{2}$ is the variance of the portfolio, $X_i$ is the percentage of the investor's assets that are allocated to the $i_{th}$ asset, and the $\sigma_{i}^{2}$ represents the variance of the asset $j$ and the covariance between the returns for assets $i$ and $j$ denoted as $\sigma _{i}{}_{j}$. Traditional asset allocation methods, such as the Markowitz theorem, reportedly provide solutions in percentages. This approach can suggest allocating half of a market share, which is often impractical~\cite{vaezi2019portfolio}. Therefore, proposing a method for determining the number of shares for each asset is crucial; this involves the conversion of expected returns, prices, and budget into interval values and determines the priority and importance of each share by framing it within a knapsack-based model.

\subsection{Knapsack Problem}
In this context, a given set of items with known sizes is selected and packed into a knapsack with a fixed capacity~\cite{kleywegt1998dynamic}. This problem is one of the simpler NP-hard problems in combinatorial optimization because it focuses on maximizing an objective function while adhering to a single resource constraint. To find the exact solutions, some techniques have been employed, such as relaxations, bounds, reductions, and other algorithmic approaches~\cite{du1998handbook}. These techniques include genetic algorithms~\cite{pradhan2014solving}, dynamic programming~\cite{schafer2021binary}, simulated annealing~\cite{delahaye2019simulated}, Tabu search~\cite{lai2019two}, and greedy algorithm~\cite{abidin2017greedy}. These classical approaches are instrumental in providing optimal solutions. Although these conventional approaches are successful and valuable, they are limitations in computation in the classical domain. Furthermore, the new paradigm of quantum computing introduces optimization techniques, including QAOA, to tackle problems such as the knapsack problem, yielding better solutions than classical computation techniques~\cite{de2019knapsack, patvardhan2015quantum}.

\subsection{Quantum Approximate Optimization Algorithm(QAOA)}
In combinatorial optimization, QAOA excels as solutions tailored for quantum computing while leveraging the strengths of classical computing~\cite{farhi2014quantum,chen2023quantum}. QAOA, a hybrid quantum-classical algorithm, has demonstrated remarkable effectiveness in addressing recent NP-hard problems, including Max-Cut~\cite{crooks2018performance}, traveling salesman problem~\cite{zawalska2022solving}, and quadratic unconstrained binary optimization (QUBO)~\cite{borle2021quantum}. Consider a combinatorial optimization problem involving an N-bit binary string represented as $z= z_1 \cdot \cdot \cdot z_N$, with a classical objective function $f(z): \{0,1\}^N \rightarrow \mathbb{R} $ is maximized. The goal is to find a solution $z$ that provides a high approximation to the maximum values of $f(z)$~\cite{ausiello1999combinatorial}. QAOA encodes this classical objective function into a quantum Hamiltonian $H_c$,
\begin{equation}
    H_{c}|z\rangle= f(z)|z\rangle. 
\end{equation}

Furthermore, $H_c$ operates diagonally on the computational basis states of  $2^N $ dimensional Hilbert space ($n$-qubit space). Ideally, the performance of the $p$-level QAOA improves with increasing $p$. For the $p$-level QAOA, the state  $\ket{+}^{\otimes N}$  is initialized, while the $H_c$ and a mixing  Hamiltonian:
\begin{equation}
\label{eq:mixing_operator}
    B=  \sum_{j=1}^N {\sigma _{j}^{x}}, 
\end{equation}
are applied alternately with controlled duration, resulting in a wave function: 
\begin{equation}
\begin{array}{l}
 \ket{\psi _{p}(\vec{\gamma } ,\vec{\beta })}  = e^{-i\beta _{p} B} e^{-i\gamma _{p} Hc} \cdots e^{-i\beta _{1} B} e^{-i\gamma _{1} H_{c}}\ket{+}^{\otimes N} .
 \end{array}
\end{equation}
This variational wave function is characterized by $2p$ variational parameters, $\gamma$ and $\beta$. The expected value of $H_c$ in this state is determined through repeated measurements on a computational basis:
\begin{equation}
    f_p(\vec{\gamma},\vec{\beta}) = \bra{\psi _{p}(\vec{\gamma } ,\vec{\beta })} H_{c}\ket{\psi _{p}(\vec{\gamma } ,\vec{\beta }  )}.
\end{equation}
Furthermore, a classical computer is used to search for the optimal parameters $(\gamma^*, \beta^*)$ and maximize the averaged output $f(\gamma^*, \beta^*)$:
\begin{equation}
\left(\overrightarrow{\gamma ^{*}} ,\overrightarrow{\beta ^{*}} \ \right) \ =\ arg\ \ \underset{\vec{\gamma } ,\vec{\beta }}{max} \ f_{p} \ (\vec{\gamma } ,\vec{\beta}).
\end{equation}
Next, the approximation is assessed by the ratio between the optimized values $f_{p}(\overrightarrow{\gamma ^{*}},\overrightarrow{\beta ^{*}})$ and the maximum possible value $f_{max} = max_zf(z)$, as the approximate ratio:  
\begin{equation}
    r\ =\ \frac{f_{p} \ \left(\overrightarrow{\gamma ^{*}} ,\overrightarrow{\beta ^{*}} \ \right)}{f_{max}}. 
\end{equation}
This approximation ratio $r$ reflects how close the QAOA solution is to the optimal classical solution. Typically, the search for this approximate ratio begins with a random initial estimate of the parameter and performing gradient-based optimization~\cite{zhou2020quantum}.  

\section{Related Work}
\label{sec:related}
Recent studies have explored hybrid approaches to enhance different quantum algorithms for knapsack-based portfolio optimization problems. For example, \cite{patvardhan2015qi} introduces Quantum-Inspired Evolutionary Algorithm (QIEAs) for difficult knapsack problems. Similarly, \cite{gunjan2024qi} used Quantum-inspired meta-heuristic approaches for constrained portfolio optimization problems. These advancements highlight the potential of hybrid quantum-classical approaches for tackling real-world optimization challenges.

Our study builds upon several key pieces of research that delve into the application of QAOA to the knapsack problem concerning portfolio optimization. QAOA has been employed as a portfolio optimization method. In this approach, the portfolio optimization problem is transformed into a binary version. Accordingly, the weight vector is discretized, with the element taking values of either 0 or 1. In~\cite{awasthi2023quantum}, a comprehensive study of quantum computing approaches for multi-knapsack problems is proposed by investigating some of the most prominent and state-of-the-art quantum algorithms using different quantum software and hardware tools. Consequently, quantum computing offers the potential for good and fast solutions to multi-knapsack optimization problems in various fields, such as logistics (allocating goods to containers), resource allocation in computing (distributing tasks among different servers), and financial portfolio optimization (allocating assets among different investment opportunities). In~\cite{brandhofer2022benchmarking}, the Markowitz models of portfolio optimization were converted into binary knapsacks. A hard constraint model was employed by incorporating hard constraints into the quantum algorithm, involving designing mixing operators based on the constraint conditions. Additionally, a combination of $XY$- and $XYY$-mixers was used to encode the constraints in the quantum circuit. $XY$-mixers were used to mix the quantum state and generate a superposition of feasible solutions, whereas $XYY$-mixers were used to enforce the hard constraints. Moreover, using a hard constraint model ensured that the quantum state evolved between feasible solutions satisfying the constraints while allowing for a high degree of flexibility in opting parameters. Another intriguing study highlights the strengths of the quantum walk optimization algorithm (QWOA)~\cite{marsh2019quantum} compared to other quantum optimization algorithms while highlighting the challenges posed by the complex and large solution space associated with its lattice structure. Moreover, the quantum mixer optimization algorithm (QMOA) was introduced as an extension of QWOA. It enhanced the efficiency of quantum optimization algorithms for portfolio optimization by reducing the number of iterations required for computations~\cite{shunza2023application}.

\section{Knapsack-based Portfolio Optimization}
\label{sec:proposed}
In this section, we explain our overall architecture, knapsack-based portfolio formulation, and QAOA for the knapsack problem.

\subsection{Overall Architecture} 
\label{sec:overall_architecture}
\begin{figure*}
    \centering
    \includegraphics[width=\linewidth]{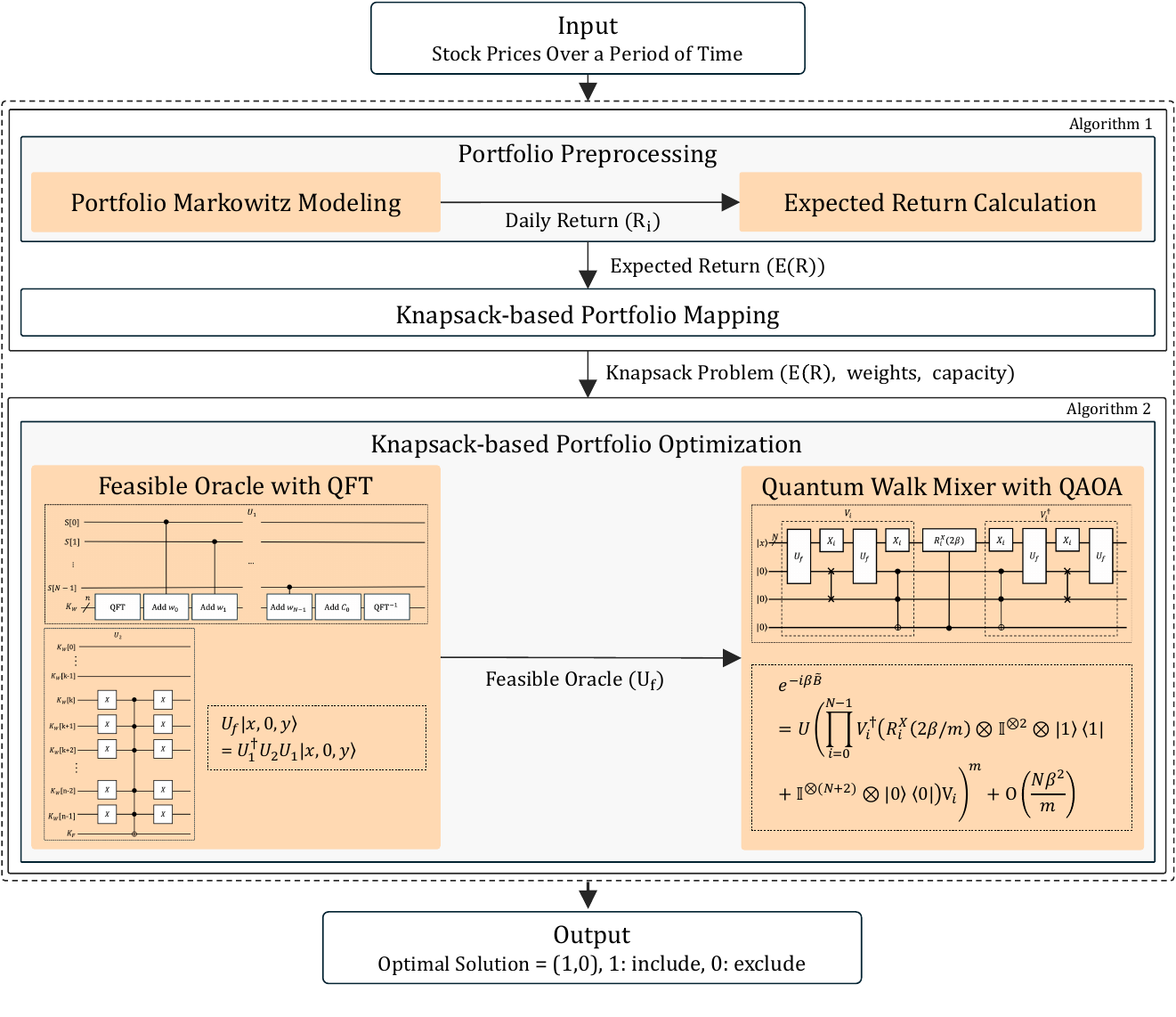}
    \caption{Workflow diagram of the quantum walk mixer with QAOA for knapsack-based portfolio optimization. The process starts with Algorithm~\ref{alg:knap_po}, where a list of stock prices is collected over time to prepare historical stock prices using the Markowitz model to calculate the expected return. This expected return is crucial in mapping the knapsack problem by considering the values, weights, and capacity of the portfolio. The optimization process in Algorithm~\ref{alg:opt_knap} involves employing a quantum walk mixer with QAOA, including the preparation of a feasible oracle. This is followed by executing the quantum walk mixer with QAOA to enhance the search for the optimal portfolio configuration. The final output is a binary string, with $1$ indicating the inclusion and $0$ indicating the exclusion of stocks from the portfolio, resulting in the optimal solution.}
    \label{fig:overall}
\end{figure*}

This section outlines the complete process from stock data preprocessing to determining the optimal portfolio configuration using a quantum-enhanced approach. As shown in Figure~\ref{fig:overall}, the workflow begins by processing historical stock prices to compute the daily returns using the Markowitz-based portfolio optimization model. Algorithm~\ref{alg:knap_po} calculates the expected returns (E(R)), which are crucial inputs for formulating the portfolio selection as a knapsack problem. This step maps the expected returns, stock weights, and capacity constraints into a knapsack framework to define the feasible portfolio configurations. In the next phase, Algorithm~\ref{alg:opt_knap} solves the knapsack-based portfolio optimization problem. Here, we first construct the feasible oracle using Quantum Fourier Transform (QFT)~\cite{weinstein2001implementation} to ensure that only valid solutions are considered for further optimization. Once a feasible solution is identified, the Quantum Walk Mixer (QWM) is applied within the Quantum Approximate Optimization Algorithm (QAOA), following the approach of ~\cite{marsh2019quantum}. The combination of QWM and QAOA (QWM-QAOA) efficiently explores the solution space, minimizing computational complexity while improving the accuracy of the optimization. The final output is a binary string, where each bit represents whether a stock is included or excluded from the portfolio. For instance, a result of (1,0) for two stocks indicates that Stock A is selected, while Stock B is excluded, thus guiding the final portfolio selection and investment strategy.

\subsection{Knapsack-based Portfolio Formulation}
\label{sec:knapsack_based_portfolio_formulation}
Our methodology is based on the fundamental principles of mean-variance optimization, focusing on the Markowitz model. This model historically maximizes the anticipated yield of a portfolio while considering a predetermined level of risk, quantified by the variance of portfolio returns. We denoted the expected return $E(R_i)$ of stock $i$ by using the mean return calculation. Next, we transformed the expected return into a knapsack problem. Subsequently, we extended this model by reinterpreting the expected return as a component of a knapsack problem, thereby aligning portfolio optimization with knapsack problem dynamics; this adaptation involves redefining the expected return of each stock as a value optimized under the constraints of total portfolio risk. This mathematical transformation effectively converts portfolio optimization into a knapsack problem:
\begin{equation}
    \begin{aligned}
        \text{maximize} \quad & R(x) = \sum_{i=1}^{n} x_i E(R_i) \\
        \text{subject to} \quad & \sum_{i=1}^{n} x_i w_i \leq C,
    \end{aligned}
\end{equation}
where the risk of including stock $i$ in the portfolio, denoted as $w_i$, and the total risk tolerance, denoted by $C$, serves as the knapsack capacity. The binary variable  $x_i$ indicates whether to include stock $i$ in the portfolio, and $E(R_i)$ represents its expected return values.

This process can be summarized as Algorithm~\ref{alg:knap_po}. It begins by extracting the historical returns of each stock within a specified time range, from the start to the end date, using the $HistoricalReturn$ function. From line $5-12$, these historical returns are then passed into the Markowitz model through the $ExpectReturn$ function, which calculates the expected returns, denoted as $E(R)$. These expected returns provide key insights into the potential profitability of the selected stocks in the portfolio problem. Specifically, the expected returns ($E(R)$) are assigned as the values in the knapsack problem, and the weights for each stock are uniformly set to 1, representing equal weighting for all assets. Afterward, the algorithm also defines the capacity constraint ($C$) of the knapsack problem, which is set to half the number of stocks, calculated as $Length(E(R))/2$. This framework effectively transforms the portfolio optimization problem into a knapsack problem formulation. This transformation allows for a more structured handling of assets and makes it possible to apply powerful combinatorial optimization techniques.
\begin{algorithm}[ht]
  \caption{Formulate Portfolio Optimization as a Binary Knapsack Problem}
  \textbf{INPUT}: $start$: start date of stock prices, $end$: end date of stock prices, $stocks$: list of stocks
  \\
  \textbf{OUTPUT}: $problem$: portfolio optimization modeled as a knapsack problem
  \label{alg:knap_po}
  \begin{algorithmic}[1]
    \Procedure{EncodeToKnapsack}{$start$, $end$, $stocks$}
      \State \texttt{// Extracte historical return}
      \State $prices \gets$ HistoricalReturn($start$, $end$, $stocks$)
      \State \texttt{// Calculate the expected return}
      \State $E(R) \gets$ ExpectReturn($prices$)
      \State \texttt{// Set a weight of 1 to each stock}
      \For{$i \gets 0$ to \texttt{Length}($E(R)$)}
        \State $weights[i] \gets 1$
      \EndFor
      \State \texttt{// Set the capacity of knapsack}
      \State $C$ = Length($E(R)$) / 2
      \State \texttt{// Formulate the problem}
      \State $problem \gets (E(R), weights, C)$
      \State \textbf{return} $problem$
    \EndProcedure
  \end{algorithmic}
\end{algorithm}

\subsection{QAOA for the knapsack problem}  
\label{sec:QAOA_knap}
After formulating the portfolio optimization as a knapsack problem, we optimized it using Algorithm~\ref{alg:opt_knap}. We employed the QWM-QAOA to prepare the mixing Hamiltonian for the QAOA process. In lines $2-10$, we set up the quantum registers and other necessary preparations for QAOA using the $QuantumRegister$ function. Line $12$ focuses on constructing the Mixing Hamiltonian $H_m$ using the QWM-QAOA approach, which is achieved via the $QWalkMixer$ function. This step requires parameters such as a predefined trotter step $m$, the number of quantum registers, weights, and ancillaries. Next, the algorithm prepares the phase Hamiltonian $H_c$ using $PhaseHamiltonian$ (line $13$), which plays a crucial role in the construction of the QAOA. More details on these concepts will be provided in the following sections. At each circuit layer $p$, the algorithm computed the expectation value (line $28$), facilitating the determination of a new set of optimal $2p$ parameter $(\gamma_{opt}, \beta_{opt})$. These optimal parameters were identified using the SHGO optimization method and implemented as $ShgoOptimizer$ line $29$. Finally, the algorithm measured probabilities (line $31$) through a measurement process. These probabilities were used to determine the optimal solution using the $OptimalChoice$ function to choose the best selection of items represented as $ar$ as the optimal choice of approximation ratio.
\begin{algorithm}
\setstretch{0.9} % Adjust to reduce vertical spacing
    \caption{Optimize Knapsack Problem Using the QWM-QAOA Algorithm}
    {\textbf{INPUT} $problem$: knapsack problem, $p$: circuit layer, $m$: trotter step count for QAOA} 
    \\
    {\textbf{OUTPUT}  $ar$: approximation ratio as optimal choice}
    \label{alg:opt_knap}
\begin{algorithmic}[1]
    \Procedure{OptimizeKnapsack}{$problem$, $l$, $m$}
    \State \texttt{// Determine problem choices} 
    \State $choices$ = Length($problem.weights$)
    \State \texttt{// Calculate the total weight} 
    \State $total\_weights$ = TotalWeight($problem.weights$)
    \State \texttt{// Setup quantum registers}
    \State $c_{reg}$ = QuantumRegister($choices$)
    \State $w_{reg}$ = QuantumRegister($total\_weights$) 
    \State \texttt{// Initialize quantum register}
    \State $a_{reg}$ = QuantumRegister($3$)
    \State \texttt{// Define quantum mixing}
    \State $H_m$= QWalkMixer($problem$,$m$, $c_{reg}$, $w_{reg}$, $a_{reg}$)
    \State \texttt{// Define phase operations}
    \State $H_c$ = PhaseHamiltonian($c_{reg}$, {$problem$)
    \State \texttt{// Start with superposition state}
    \State $|\psi_0\rangle$ = $\ket{+}^{\otimes choices}$  
    \For{$ k = 1$ to $p$}
        \If{$k \gets 1$ }
            \State $\gamma_k,\beta_k$ = Initialize()
        \Else 
             \State ($\gamma_k$,$\beta_k$) = ($\gamma_{opt}, \beta_{opt}$)
         \EndIf
            \State \texttt{// Apply the unitary operators}
            \State $U_B(\beta_k)$ = $e^{-i \beta_k H_M}$
            \State $U_c(\gamma_k)$ = $e^{-i \gamma_k H_c}$
            \State $\ket{\psi_k} = U_c(\gamma_k) U_B(\beta_k)  \ket{\psi_{k-1}}$
            \State \texttt{// Define the QAOA cost function}
            \State {$cost(\gamma,\beta)$=$\bra{\psi_k}H_c\ket{\psi_k}$}
            \State \texttt{// Optimize the cost function}
            \State ($\gamma_{opt}, \beta_{opt}$)= ShgoOptimizer($cost(\gamma,\beta)$)  
    \EndFor
    \State \texttt{// Measure the final state}
    \State $probabilities$ =  MeasureProbabilities($\ket{\psi_p}$
    \State \texttt{// Obtain the optimal probability}
    \State $ar$ =  GetOptimalSolution($probabilities$)
    \State {\textbf{return} $ar$}
    \EndProcedure}
\end{algorithmic}
\end{algorithm}

\subsubsection{Feasibility Oracle} 
\label{sec:fease_oracle}
The feasibility oracle is used as a hypothetical subroutine that instantly determines whether a proposed solution to the knapsack problem violates any constraints~\cite{chawla2023multi}. We can explore solutions within a well-defined space of bitstrings by representing our portfolio model as a binary knapsack problem. We defined $K(N) = (0,1)^N$ as the set of all possible bitstrings of length $N$ representing potential portfolio choices. Furthermore, each possible choice of any of the $N$ items is represented by a bitstring $x \in K(N)$. ). Thus, the subset feasible solution was denoted as $F$ for the knapsack problem, and the feasibility function was defined as  
\begin{equation}
\label{eq:feasible_func}
    f: K(N) \to \{0, 1\}, \quad x \mapsto f(x) = 
    \begin{cases}
    1, & \text{if } x \in F, \\
    0, & \text{otherwise}.
    \end{cases}
\end{equation}
Considerably, the feasibility oracle to be unitary $U_f$ as:
\begin{equation}
    \label{eq:f_oracle}
    U_f \ket{x, y} = \ket{x, y \oplus w(x)}, \quad \forall x \in K(N), \quad y \in K(1)
\end{equation}
In portfolio optimization, a state $\ket{x}$ symbolizes a specific stock allocation, which is deemed feasible (i.e., $x \in F$) if the total weight $w(x)$ does not exceed the capacity $C$. This oracle toggles a flag qubit $\ket{y}$ based on the feasibility of the state $\ket{x}$, representing a possible portfolio configuration.

Next, we allocated qubits for storage as follows: $S$ to record the formulated knapsack choices, $K_w$ to hold the weight of the item choice, and $K_F$ as the flag qubit indicating the feasibility of the state $S$. Remarkably, the number of qubits required for $S$, $K_w$, and $K_F$ are deonted by $Q_S$, $Q_w$, and $Q_F$ respectively. Upon that, the total number of qubits that are required is $Q_S + Q_w + Q_F$. 

Furthermore, the total weight $w(x)$  is calculated by adding the weight of each item to register $K_w$, controlled by the corresponding bit in register $S$. We also compare the computed weight $w(x)$ to the capacity $C$ using an inequality check facilitated by a multiple-controlled NOT gate.

For suitable $W_0 \in \mathbb{N}$, the inequality can be verified using the following condition:
\begin{equation}
    \label{eq:constraint_u}
    w(x) \leq C \Leftrightarrow w(x) + C_0 < C + C_0 + 1
\end{equation}
This condition ensures that the binary representation of $w(x)+C_0$ has zeros in all positions beyond $k$.
The feasibility oracle $U_f$ employs two primary unitary operations-$U_1$ and $U_2$. $U_1$ augments $K_w$ with a predetermined offset $C_0$, facilitating the binary representation required for the feasibility check; $U_2$ is applied conditionally based on $U_1$'s outcome.

We begin by illustrating how $U_1$ is prepared. Quantum Fourier Transform (QFT)~\cite{weinstein2001implementation} is employed within the unitary operation $U_1$ to facilitate the process of adding weights in the knapsack problem. 

\begin{figure}[ht]
    \centering
    \includegraphics[width=85mm]{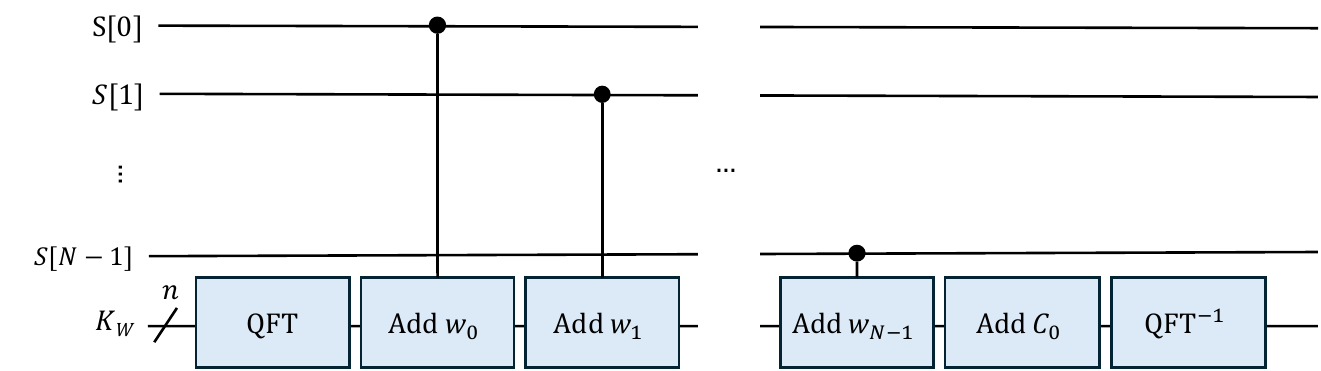}
    \caption{Circuit implementation of $U_1$. It encodes the total weights of item selections into the ancillary register $K_W$ using quantum Fourier transform-based arithmetic. A predefined offset $C_0$ is added to the register to facilitate subsequent feasibility checking.}
    \label{fig:u1}
\end{figure}

\begin{figure}[ht]
    \centering
    \includegraphics[width=55mm]{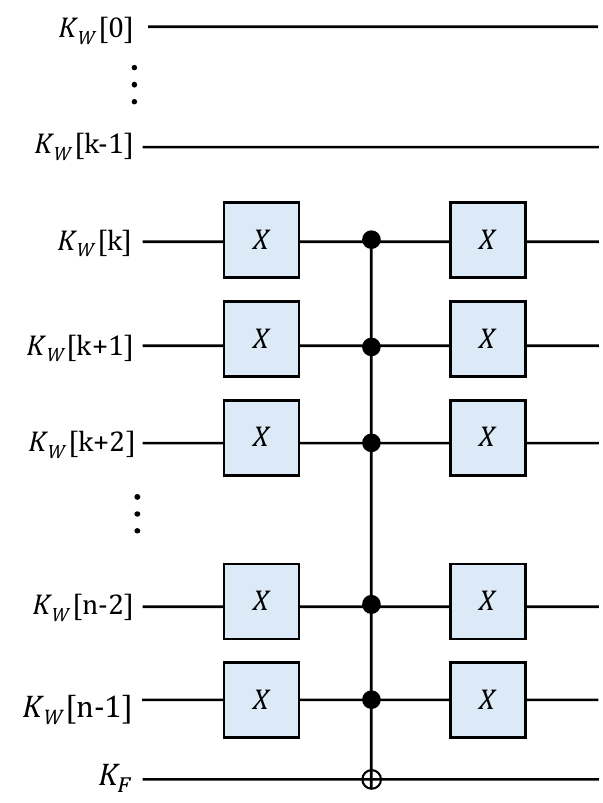}
    \caption{Circuit implementation of $U_2$. It executes a conditional check based on the total weight calculated by $U_1$ using several multi-controlled NOT gates to flip the ancillary qubit if the weight condition is satisfied.}
    \label{fig:u2}
\end{figure}

This process starts with the initialization of the ancillary weight register $K_w$ to $\ket{0}$. Next, QFT is applied to $K_w$ to prepare the register for quantum addition, even though the QFT does not change the state for state $\ket{0}$. This step is necessary for nonzero initial states $K_w$. Post-state transformation of $K_w$ from the computational basis to the Fourier basis, the weight representation was distributed across the amplitude of the quantum state:
\begin{equation}
    QFT\ket{K_w} = \frac{1}{\sqrt{2^n}} \sum_{k=0}^{2^n-1} e^{2\pi i \cdot 0 \cdot k / 2^n} \ket{k}
\end{equation}
where we use the notaion $e(t) = exp(2\pi it)$, and $k \in \mathbb{Z}$ is the binary representation of $k_n-1,\cdot\cdot\cdot,k_0$. Furthermore, weights were added to $ QFT\ket{K_w}$ using controlled operations applying phase rotations. These operations are conditioned on the bits of the bitstring $x$ indicating the selection of items. Each bit in $x$ determines whether the corresponding weight $w_i$ should be added to $K_w$. The addition in the Fourier space was related to phase rotation and implemented by a sequence of controlled phase gates. The magnitude of the rotation was determined by $w_i$ and the position of the bit controlling the operation. The implementation of controlled additions can be described as follows: for each selected item (where the corresponding bit in $x$ is $\ket{1}$, a controlled phase rotation is applied to $K_w$. The phase added to each computational basis state $\ket{k}$ within $K_w$ is proportional to $w_i$:
\begin{equation}
    \ket{K_w} = 
    \begin{cases}
        \ket{K_w + w_i}, & \text{if } x_i = 1, \\
        \ket{K_w}, & \text{otherwise.}
    \end{cases}
\end{equation}
This phase rotation effectively encodes the addition of $w_i$ into the quantum state. After completing this process, the step transformed the register back to the computational basis required, where
\begin{equation}
    QFT^{-1} \ket{K_w}
\end{equation}
The inverse QFT decodes the phase information back into a computational state representing the total weight of the selected items. If QFT encodes the weight as a superposition of phases, the inverse QFT converts these phases back into a binary number the sum of the weights. The final state of the register $K_w$ after applying $U_1$ was the binary representation of the total weight of the items:
\begin{equation}
    U_1 \ket{x, 0, y} = \ket{x, w(x) + C_0, y}
\end{equation}
Figure~\ref{fig:u1} shows how $U_1$ is implemented using an algorithm based on QFT. 
Following the implementation of the unitary operation  $U_1$, $U_2$ is described in Figure~\ref{fig:u2}, which comprises a quantum circuit that conditionally modifies the state of an ancillary qubit concerning the sum of weights represented in the register $K_w$. Referring to Equation~\ref{eq:constraint_u}, the ancillary qubit  $\ket{y}$ underwent a state flip to signal a valid configuration when the total weight $w(x)$, augmented by a constant offset $C_0$ is less than the predefined threshold $C +C_0 +1$. This transformation was executed through several multi-controlled-NOT gates, where each gate was influenced by a qubit distinct from the weight register. The successful transition of the ancillary qubit’s state post $U_2$ signified a feasible solution, adhering to the knapsack’s capacity, which effectively segregates the solution space into permissible and impermissible weights.

\subsubsection{Quantum Walk Mixer for Enforcing Constraints}
\label{sec:quantum_walk_mizer_for_enforcing_constraints}
This process relies on the mixing operator $U_B (\beta)$ generated by mixing Hamiltonian $B$ from Equation~\ref{eq:mixing_operator} as follows:
\begin{equation}
    U_B(\beta) = e^{-i \beta B}
\end{equation}
Alternatively, by leveraging the feasible function $f$ in Equation~\ref{eq:feasible_func} and neighboring states $n_i(x)$ for exploring the solution space,  $B$ can be described as
\begin{equation}
\label{eq:alter_mixing}
    B\ket{x} = \sum_{i=0}^{N-1} f_i(x) \ket{n_i(x)}, \quad \forall x \in K(N)
\end{equation}
where  $n_i(x)$ is the $i_{th}$ neighbor of $x$ ($x$ with $i_{th}$ bit flipped). Using this alternative representation, it is easy to observe that $\forall x, x' \in K(N)$.
\begin{equation}
    \label{eq:hamil_qw}
    \bra{x} B \ket{x'} = 
    \begin{cases}
        1, & \text{if } \operatorname{Ham}(x, x') = 1, \\
        0, & \text{otherwise}
    \end{cases}
\end{equation}
where Ham $(x, x')$ represents the Hamming distance between two binary strings $x$ and $x'$, which is the count of positions where the corresponding bits differ. The exact implementation of the desired mixing operator $U_B(\beta)$ presents challenges due to its Hamiltonian, requiring additional resources and non-commuting elements. To address this, we employ an alternative operator $\tilde{B}$, closely resembling $B$. This operator is constructed from $V_i$, and the inverse $V_i^\dagger$ for encoding feasibility information of neighboring states into auxiliary qubits that enable the controlled state manipulation. Additionally, the operator includes single-qubit $X_i$ gates for creating neighboring state exploration. Moreover, $U_f$ the feasible oracle is included to determine the feasibility of states by ensuring that only valid states are mixed. Finally, $R_i^X(2\beta)$ aids in adjusting the amplitudes of states based on feasibility and further refining the mixing process, whereas auxiliary qubits are employed to store feasibility information.

\begin{figure}[ht]
    \centering
    \includegraphics[width=85mm]{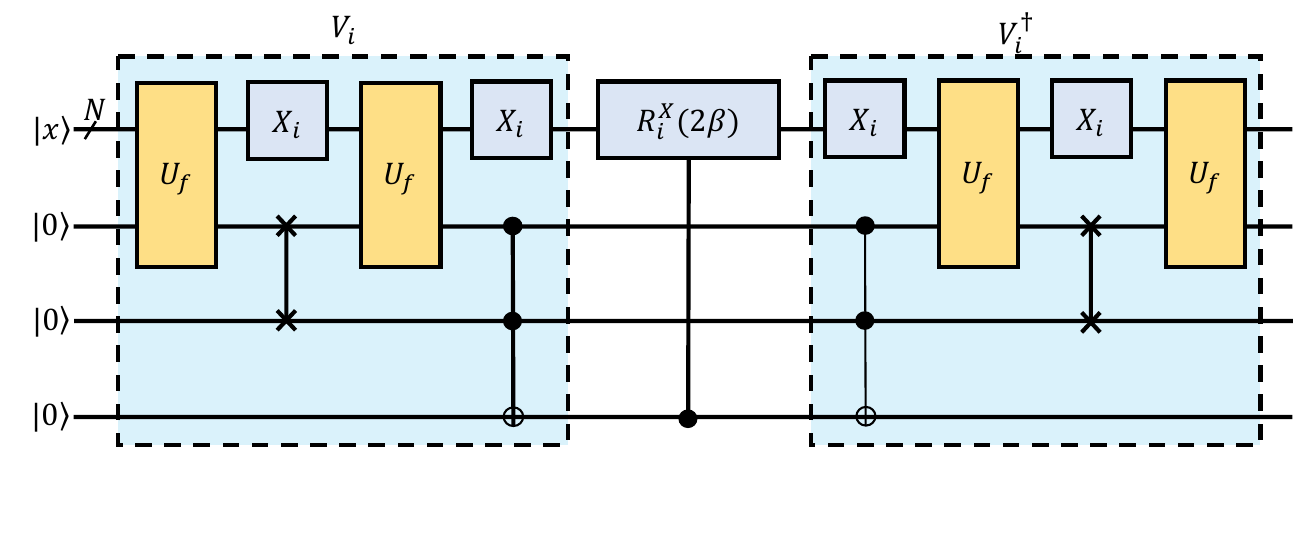}
    \caption{General circuit for a single-qubit quantum walk mixer. It illustrates the core operations involved in mixing a quantum state with its feasible neighbor, guided by feasibility oracles and single-qubit rotations, as part of a constrained optimization algorithm.}
    \label{fig:qwqaoa}
\end{figure}
As shown in Figure~\ref{fig:qwqaoa}, the process starts from:
\begin{equation}
\label{eq:unitary_qw}
    V_i V_i^{\dagger} \ket{x,0,0,0} = \ket{x,0,0,0}
\end{equation}
which implies that the behavior of $V_i$ when acting on a state $\ket{x,0,0,0}$ is to transform it into a new state with additional information encoded in the auxiliary qubits, dependent on the feasibility function $f$ and its value at neighboring points. Therefore, the state of $V_j$ can be described as
\begin{equation}
\label{eq:vj_check}
    V_i \ket{x, 0, 0, 0} = \ket{x, f(n_i(x)), f(x), f_i(x)}
\end{equation}
Next, neighboring state mapping is performed using $V_i ^\dagger$ to determine whether the original and neighboring states have the same feasibility. Thus, Equation~\ref{eq:unitary_qw} can be written as

\begin{equation}
\begin{aligned}
    &V_i^{\dagger} \ket{x, f(n_i(x)), f(x), f_i(x)} \\
    &\quad = \ket{n_i(x), f(x) \oplus f(n_i(x)), f(x) \oplus f(n_i(x)), f_i(x) \oplus f_i(x)} \\
    &\quad = \begin{cases}
        \ket{n_i(x), 0, 0, 0}, & \text{if } f(x) = f(n_i(x)), \\
        \ket{n_i(x), 1, 1, 0}, & \text{otherwise}
    \end{cases}
\end{aligned}
\label{eq:bit_flip}
\end{equation}
Then, the bit flip of the $i-{th}$ qubit acts as:
\begin{equation}
    X_i \ket{x} = \ket{n_i(x)}
\end{equation}
Therefore, we can describe the encoded feasibility information after applying $V_i$ and based on Equation~\ref{eq:vj_check} as follows:
\begin{equation}
\begin{aligned}
    & (X_i \otimes \mathbb{I}^{\otimes 2} \otimes \ket{1}\bra{1}) V_i \ket{x, 0, 0, 0} \\
    & \quad = f_i(x) \ket{n_i(x), f(n_i(x)), f(x), f_i(x)}
\end{aligned}
\label{eq:operator_action}
\end{equation}
Next,
\begin{equation}
\begin{aligned}
    & V_i^{\dagger} \left( X_i \otimes \mathbb{I}^{\otimes 2} \otimes \ket{1}\bra{1} \right) V_i \ket{x, 0, 0, 0} \\ 
    & \quad = f_i(x) \ket{n_i(x), 0, 0, 0}
\end{aligned}
\end{equation}
Thus, for a given $ \tilde{B}$ operator, which approximates the original $B$ operator, we have
\begin{equation}
    \tilde{B} = \sum_{i=0}^{N-1} V_i^{\dagger} \left( X_i \otimes \mathbb{I}^{\otimes 2} \otimes \ket{1}\bra{1} \right) V_i
\end{equation}
Moreover, $\tilde{B}$ implements $B$ given three ancillary qubits, or more precisely, it is represented as:
\begin{equation}
\begin{aligned}
    \tilde{B} \ket{x, 0, 0, 0} &= \sum_{i=0}^{N-1} f_i(x) \ket{n_i(x), 0, 0, 0} \\
    &= (B \otimes \mathbb{I}^{\otimes 3}) \ket{x, 0, 0, 0}, \quad \forall x \in K(N).
\end{aligned}
\end{equation}
It follows that
\begin{equation}
\label{eq:ub}
\begin{aligned}
    e^{-i\beta \tilde{B}} \ket{x, 0, 0, 0} &= e^{-i\beta (B \otimes \mathbb{I}^{\otimes 3})} \ket{x, 0, 0, 0} \\ 
    &= \left( U_B(\beta) \otimes \mathbb{I}^{\otimes 3} \right) \ket{x, 0, 0, 0}
\end{aligned}
\end{equation}
where $\tilde{B}$ is used to generate  $U_B$. Moreover,  $\tilde{B}$ is not commute and should use three ancilla qubits. Therefore, the approximate implementation of  $e^{-i \beta \tilde{B}}$ needs to use the trotter product formula. Meanwhile, using the identity $R_i^X(2\beta)$, the state representing the rest of the circuit in Figure~\ref{fig:qwqaoa}.
\begin{equation}
\begin{aligned}
    & e^{-i\beta V_i^{\dagger} \left( X_i \otimes \mathbb{I}^{\otimes 2} \otimes \ket{1}\bra{1} \right) V_i} \\
    & = V_i^{\dagger} \left( R_i^X(2\beta) \otimes \mathbb{I}^{\otimes 2} \otimes \ket{1}\bra{1} + \mathbb{I}^{\otimes (N+2)} \otimes \ket{0}\bra{0} \right) V_i
\end{aligned}
\label{eq:transformation}
\end{equation}
where $R_{i}^{X}( 2\beta ) = cos(\beta)\mathbb{I}^{\otimes N} - i sin(\beta)X_i$. Furthermore, we optimized $U_{B}( \beta ) \otimes \mathbb{I}^{\otimes 3}$ with the trotter product formula as
\begin{equation}
\begin{aligned}
    e^{-\beta\tilde{B}} = & \left(\prod_{i=0}^{N-1} V_j^{\dagger} \left( R_i^X \left(\frac{2\beta}{m}\right) \otimes \mathbb{I}^{\otimes 2} \otimes \ket{1}\bra{1} \right.\right. \\
    & \left.\left. \quad + \mathbb{I}^{\otimes (N+2)} \otimes \ket{0}\bra{0} \right) V_j \right)^m \\ 
    & \quad + O\left(\frac{N\beta^2}{m}\right)
\end{aligned}
\label{eq:expansion}
\end{equation}
Moreover, to maintain $\beta$ the validity of the approximation within the new parameter range, the $\beta$ range needs to be adjusted accordingly to $\beta \in [0,m\pi)$.

As an objective function, we use the value function  $v$. Therefore, the corresponding phase separation can be described as follows:
\begin{equation}
    U_{C}(\gamma) \ket{x} = e^{-i\gamma v(x)} \ket{x}
\end{equation}

\subsubsection{Number of Qubit Requirements} 
\label{sec:number_of_qubit_requirements}
The presented illustration implies that using a feasibility oracle with $n_f$ ancilla qubits requires $N + n_f + 3$ qubits. In contrast, there is an exigency for an additional three qubits, specifically for generating the unitary $U_B$ (Equation~\ref{eq:ub}).

\section{Evaluation}
\label{sec:evaluation}
This section presents the numerical results by evaluating the effectiveness of the proposed algorithm across different asset numbers along with the impact of parameter configuration.

\subsection{Experimental Setup}
\label{sec:experimental_setup}
We present the experimental setup for evaluation, specifically the datasets, hyperparameter configuration, and model setup used to evaluate our proposed approach.

\subsubsection{Hyperparameter Configuration}
\label{sec:hyperparameter}
Initially, we implemented the proposed method using the Qiskit library~\cite{aleksandrowicz2019qiskit}, a widely used open-source quantum computing framework. We run the algorithms on the QASM simulator, a tool provided by Qiskit for simulating quantum circuits, which helped us gain insights into the behavior and performance of the algorithms before moving on to the real quantum device. Moreover, we used a fake backend noise from 127 qubits of IBM FakeWashington. While using the real device, we used IBM Nazca, a 127-qubit quantum device, for quantum computations, which offers the opportunity to test the algorithms in a real-world quantum computing environment. The experimental phase involves conducting several tests in which we select varying stocks from $yahoo finance$, such as Apple Inc. (AAPL), Amazon.com Inc. (AMZN), Alphabet Inc. (GOOGL), Microsoft Corporation (MSFT), NVIDIA Corporation (NVDA), and Tesla, Inc. (TSLA), Netflix, Inc. (NFLX), and Visa Inc. (N). Notably, the selection of these stocks was based on different scenarios, including all cases from two to eight stocks. Postselection, we proceeded with the experimental setup of the proposed method. During the optimization process, we focused on optimizing QAOA, which was achieved by optimizing the $2p$ angle parameters, i.e.,  $\beta$ and $\gamma$, using the classical optimizer SHGO~\cite{endres2018simplicial}. Additionally, we integrated quantum walk to boost optimization by leveraging its benefits for refining the process and achieving superior results. 
\begin{table}[ht]
\centering
\caption{Hyperparameter Configuration}
\begin{tabular}{lc}
    \hline
    \textbf{Hyperparameter}                   & \textbf{Value} \\ \hline
    Number of circuit layers                  & 3              \\ 
    Number of trotter steps                   & 3              \\ 
    Financial data provider                   & yahoo finance       \\ 
    Portfolio valuation module                & PyPortfolioOpt \\ 
    Classical optimizer                       & SHGO           \\ 
    Framework for architecture implementation & Qiskit         \\ \hline
\end{tabular}
\label{tab:hyperparams}
\end{table}

\subsubsection{Problem Setup}
\label{sec:evaluation_metric}
To evaluate our proposed method's performance, we defined our algorithm's problem based on the number of assets to be optimized (Table~\ref{tab:configuration}). First, we selected 2–8 subsets of stocks using the capabilities of PyPortfoliopt~\cite{Martin2021}. The values represent the expected return of stocks counting from January 1, 2018, to January 1, 2023. The knapsack capacity $Cap$ is the maximum weight, where in our case we set it to be half of the total number of assets. This step clarifies the problem and enables us to tailor our algorithm accordingly. We aimed to evaluate the effectiveness of the QAOA optimization technique in portfolio optimization. To achieve this, we specifically evaluated the performance of the algorithm using the best-known solution (BKS), a result of using a classical algorithm~\cite{martello1999dynamic}, where $1$ signifies selection and $0$ denotes exclusion. This technique compared the results obtained from our algorithm with those of the classical solution, providing a benchmark for assessing its performance and effectiveness. By undertaking this evaluation, we can gain valuable insights into the capabilities and limitations of the QAOA optimization technique in portfolio optimization.  

\setlength{\tabcolsep}{4pt} % Reduce column separation
\renewcommand{\arraystretch}{0.9} % Reduce row height
\begin{table}
\centering
\caption{The optimal stock selection and values for various portfolio sizes under the binary knapsack constraint. It includes the number of selected stocks (\#Ss), stock types (Type), their values, knapsack capacity (Cap), Best Known Solution (BKS), and qubits used (\#Qs). The BKS column shows the binary string representation of the optimal solution.}
\label{tab:configuration}
\begin{tabular}{cccccc}
\hline
\textbf{\#Ss} & \multicolumn{1}{c}{\textbf{Type}} & \multicolumn{1}{c}{\textbf{Values}} & \textbf{Cap} & \textbf{BKS} & \textbf{\#Qs} \\ 
\hline
\multirow{2}{*}{2} & MSFT & 0.2430 & \multirow{2}{*}{1} & \multirow{2}{*}{(0,1)} & \multirow{2}{*}{7} \\ 
                   & AAPL & 0.2602  &                    &                        &                    \\ \hline
\multirow{3}{*}{3} & MSFT & 0.2430  & \multirow{3}{*}{1} & \multirow{3}{*}{(0,1,0)} & \multirow{3}{*}{8} \\ 
                   & AAPL & 0.2602  &                    &                        &                    \\ 
                   & NVDA & 0.2430  &                    &                        &                    \\ \hline
\multirow{4}{*}{4} & MSFT & 0.2430  & \multirow{4}{*}{2} & \multirow{4}{*}{(1,1,0,0)} & \multirow{4}{*}{10} \\ 
                   & AAPL & 0.2602  &                   &                        &                    \\ 
                   & GOOGL & 0.1047 &                    &                        &                    \\ 
                   & NVDA & 0.2430  &                    &                        &                    \\ \hline
\multirow{5}{*}{5} & MSFT & 0.2430  & \multirow{5}{*}{2} & \multirow{5}{*}{(0,1,0,0,1)} & \multirow{5}{*}{11} \\ 
                   & AAPL & 0.2602  &                    &                        &                    \\ 
                   & GOOGL& 0.1047 &                   &                        &                    \\ 
                   & AMZN & 0.0716  &                    &                        &                    \\ 
                   & NVDA & 0.2430  &                   &                        &                    \\ \hline 
\multirow{6}{*}{6} & MSFT & 0.2430  & \multirow{6}{*}{3} & \multirow{6}{*}{(0,1,0,0,1,1)} & \multirow{6}{*}{12} \\ 
                   & AAPL & 0.2602  &                    &                        &                    \\ 
                   & GOOGL& 0.1047 &                   &                        &                    \\ 
                   & AMZN & 0.0716  &                    &                        &                    \\ 
                   & NVDA & 0.2430  &                   &                        &                    \\
                   &   TSLA  & 0.4203      &                    &                        &                    \\ \hline
\multirow{7}{*}{7} & MSFT &0.2430  & \multirow{7}{*}{3} & \multirow{7}{*}{(0,1,0,0,1,1,0)} & \multirow{7}{*}{13} \\ 
                   & AAPL &  0.1870  &                    &                        &                    \\ 
                   & GOOGL &  0.1749  &                    &                        &                    \\ 
                   & AMZN &  0.1898   &                    &                        &                    \\ 
                   & NVDA &  0.2856   &                    &                        &                    \\ 
                     &   TSLA  & 0.4203       &                    &                        &                    \\
                   & NFLX    & 0.0797       &                    &                        &                    \\ \hline
\multirow{8}{*}{8} & MSFT & 0.2430 & \multirow{8}{*}{4} & \multirow{8}{*}{(1,1,0,0,1,1,0,0)} & \multirow{8}{*}{15} \\ 
                   & AAPL &0.1899  &                    &                        &                    \\ 
                   & GOOGL &  0.1780  &                    &                        &                    \\ 
                   & AMZN &   0.1903   &                    &                        &                    \\ 
                   & NVDA &  0.2874  &                    &                        &                    \\ 
                    &   TSLA  &  0.4203       &                    &                        &                    \\ 
                   & NFLX     & 0.0797      &                    &                        &                    \\
                   &  V    & 0.1341      &                    &                        &                    \\ \hline 
\end{tabular}
\end{table}

\subsection{Performance Evaluation}
\label{sec:performance_evaluation}
In this section, we present our model performance and its sensitivity analysis. 
\subsubsection{Model Performance}
\label{sec:model_performance}
\begin{figure}[ht]
    \centering 
    \includegraphics[width=85mm]{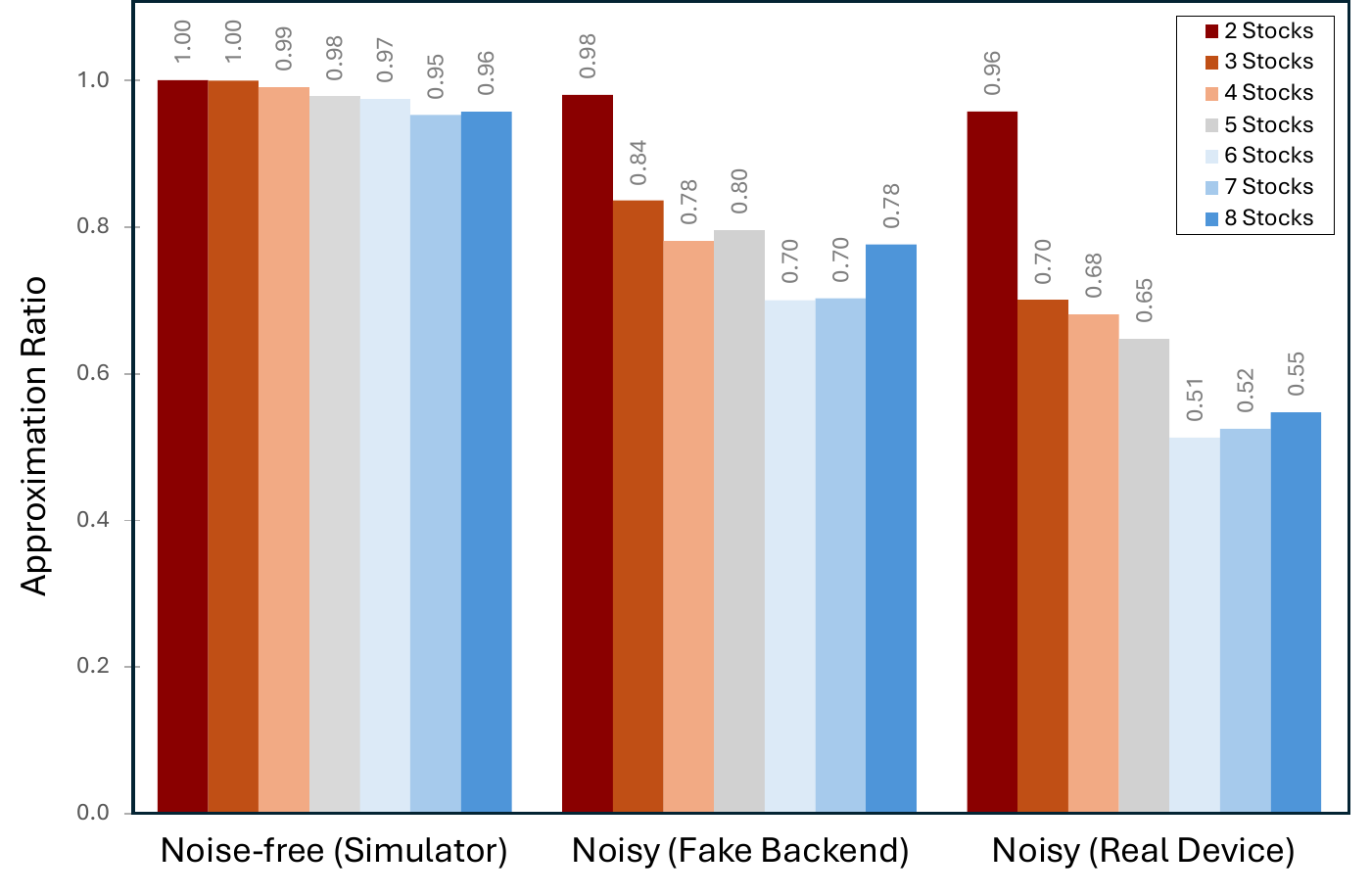}
    \caption{Approximation ratio for different scenarios involving various numbers of stocks under three distinct environmental conditions: noise-free with the simulator, noisy with Fake Washington, and noisy with the IBM Nazca device.}
    \label{fig:ratio}        
\end{figure}

By meticulously exploring the variable settings within our QWM-QAOA, we identified that a circuit layer depth of $p = 3$ and a Trotter step value of $m = 3$ yielded the best results. Figure~\ref{fig:ratio} illustrates the approximation ratios for knapsack-based portfolio optimization with varying numbers of stocks (ranging from 2 to 8 stocks) under three different environmental conditions: noise-free (simulator), noisy (using FakeWashington as a simulated backend), and noisy (real device: IBM Nazca with 127 qubits).
\begin{itemize}
    \item Noise-free (Simulator): The approximation ratios remain very high across all stock configurations, ranging from 100\% (for 2 and 3 stocks) to 96\% (8 stocks), indicating that the algorithm performs near-perfectly in an ideal, noise-free environment.
    \item Noisy (Fake Backend): The introduction of noise through the simulated FakeWashington backend results in a decline in performance. The approximation ratios range from 98\% to 70\%, with lower values observed for larger portfolios. For example, 2 stocks achieve an approximation ratio of 98\%, while 6 and 7 stocks drop to 70\%. Interestingly, for 8 stocks, the approximation ratio slightly improves to 78\%, indicating a non-linear impact of noise for larger portfolios.
    \item Noisy (Real Device): The experiment is executed on the real IBM Nazca device with the approximation ratios dropping further, ranging from 96\% (for 2 stocks) to 51\% (for 6 stocks). Surprisingly, for 7 and 8 stocks, the approximation ratios increased to 52\% and 55\%, respectively.
\end{itemize}
This performance reduction is due to real-world factors, such as gate errors and decoherence, which exert a more significant impact as the number of stocks increases. This unexpected result hints at the potential for further improvements, even as the number of stocks increases in a noisy environment~\cite{johnstun2021understanding}. These issues highlight the importance of advanced error mitigation techniques, though such methods are beyond the scope of this study~\cite{alam2019analysis, gaur2023nr}.

\subsubsection{Sensitivity Analysis}
\label{sec:sensitivity_analysis}
We conducted a sensitivity analysis of key parameters using our QWM–QAOA algorithms, such as the circuit layer $(p)$ and the number of trotter steps $(m)$. These parameters are crucial for shaping the effectiveness of our proposed algorithm. Through a methodical examination and evaluation, we present details of the relationship between these parameters and how much they influence the overall efficiency and resilience of our QWM–QAOA.
\begin{itemize}
    \item \textbf{Circuit Layer:} Figure~\ref{fig:depth} shows that as the circuit layer increases beyond the initial value of $p\geqslant3$, the approximation ratio of the diverse portfolio optimization result consistently reaches an impressive result ranging from 100\% to 95\%, categorically based on the setup cases of portfolio optimization. This empirical evidence emphatically establishes that our performance consistently produces an efficient result even at some circuit layers.
    \begin{figure}[ht]
        \centering
        \includegraphics[width=85mm]{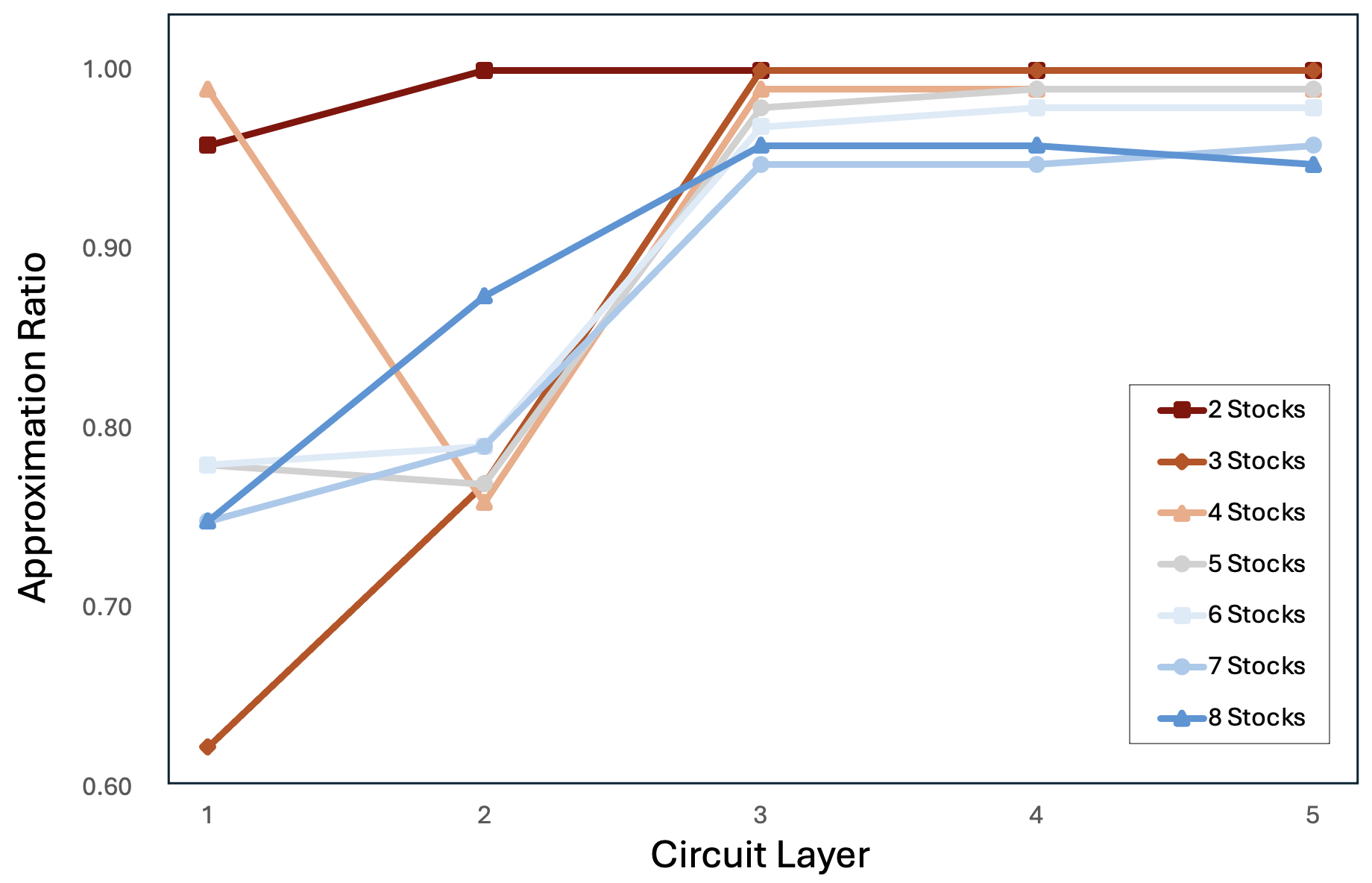}
        \caption{Approximation ratio of the circuit layer $(p)$ is 1–5 across all problems. Remarkably, when $p\geqslant3$, the approximation ratio reaches 100\%, indicating that the proposed method is efficient even at the lower circuit layer.}
        \label{fig:depth}
    \end{figure}

    \item \textbf{Trotter Step in QWM-QAOA:} We aimed to find the parameter $m$, representing the trotter step counts in the QWM–QAOA algorithm while maintaining a constant circuit layer of $p=3$ and optimize $\beta$, $\gamma$ for the resulting circuit of 1–5. Table~\ref{tab:penalty} shows that the approximation ratio for various stock selections increasingly exhibits a better result starting from $m=3$, ranging from 100\% to 95\%. When $m=5$, we obtained efficient optimization results from our proposed method ranging from 100\% to 96\%. The approximation ratios demonstrate how increasing the trotter steps improves the portfolio optimization results, particularly when $m\geqslant3$, yielding consistently higher performance across different portfolio problem sizes.
    \begin{table}[ht]
        \centering
        \caption{Performance of QWM-QAOA with varying Trotter steps across different stock selection scenarios. The table presents the approximation ratios for various numbers of stocks, ranging from 2 to 8, using the QWM-QAOA algorithm with different Trotter step values (1 to 5).}
        \begin{tabular}{c|c|c|c|c|c}
        \hline
        \textbf{\#Stocks} & \textbf{\boldmath{$m$} = 1} & \textbf{\boldmath{$m$} = 2} & \textbf{\boldmath{$m$} = 3} & \textbf{\boldmath{$m$} = 4} & \textbf{\boldmath{$m$} = 5} \\ \hline 
        2                  &  1.00            & 0.96             & 1.00             & 1.00             & 1.00             \\ 
        3                  &  1.00            & 0.60             & 1.00             & 1.00             & 1.00             \\ 
        4                  &  0.77            & 0.99             & 0.99             & 0.99             & 0.99             \\ 
        5                  &  0.96            & 0.96             & 0.98             & 0.98             & 0.98             \\ 
        6                  &  0.96            & 0.97             & 0.98             & 0.98             & 0.98             \\
        7                  &  0.96            & 0.96             & 0.95             & 0.96             & 0.96             \\
        8                  &  0.96            & 0.96             & 0.96             & 0.96             & 0.96             \\\hline
    \end{tabular} 
    \label{tab:penalty}
    \end{table}
\end{itemize}

\section{Discussion and Limitation}
\label{sec:discussion}
our research provides a novel application of quantum computing, specifically through the Quantum Walk Mixer combined with a Quantum Approximate Optimization Algorithm (QWM-QAOA) integrating with Quantum Fourier Transform (QFT), to address the knapsack-based financial portfolio optimization problem. By transforming the traditional portfolio optimization problem into a binary knapsack problem formulation, we were able to leverage quantum computing to provide efficient and effective optimization strategies, particularly for NP-hard problems. The results from our experiment on noise-free simulation indicate that the QWM-QAOA approach can yield high approximation ratios in both fake backend and real quantum devices, demonstrating the potential of quantum computing in financial optimization tasks. Our study also highlights the importance of various QAOA parameters, such as circuit layer $p$ and trotter step $m$, in optimizing the performance of our proposed algorithm. Through sensitivity analysis, we found that a circuit layer of $p\geqslant3$ and the number of trotter steps $m\geqslant3$ provided the most converged results, with approximation ratios consistently reaching between 100\% and 96\% in the best cases. This empirical evidence supports the notion that QAOA with the proper configuration, can outperform classical optimization approaches in the context of financial portfolio optimization problems. However, our experiments on real quantum hardware revealed some limitations. The approximation ratio dropped significantly due to unavoidable hardware noise such as gate errors and decoherence errors, with real-device performance yielding ratios as low as 50\%. Despite these current limitations of quantum hardware, 
this finding emphasizes the pressing need for continued research in quantum error mitigation (QEM) techniques and hardware advancements to unlock the full potential of our proposed QWM-QAOA method on portfolio optimization problems in real-world applications. 

\textbf{Noise and Hardware Constraints:} The most significant limitation encountered in this study was the impact of overall gate errors when using real quantum devices~\cite{satpathy2023analysis}. While the QAOA algorithm performed well in simulators and fake backend environments, real hardware results showed a substantial drop in approximation ratios (below 50\%). This limitation reflects the current state of NISQ devices, which are not yet fully capable of handling complex optimization problems without significant fidelity loss.

\textbf{Scalability:} While our approach worked well for smaller portfolios, scaling QWM-QAOA to larger datasets faces challenges due to the increasing number of required qubits and the limitations of current quantum hardware, such as limited qubit availability and short coherence times. These constraints hinder its applicability to larger portfolio optimization problems. Advancements in quantum hardware and quantum error mitigation (QEM) techniques along with hybrid approaches are needed to extend QWM-QAOA to more complex real-world scenarios.

\textbf{Limited Dataset:} The experimental evaluations were conducted on a small selection of stocks, ranging from two to eight, providing a controlled setting to demonstrate the algorithm’s effectiveness. However, to determine its practical value, future evaluations must incorporate historical data, transaction costs, and market dynamics. Expanding the tests to larger and more diverse datasets will be crucial for assessing the broader applicability of the QWM-QAOA approach in more complex, real-world financial scenarios.

\section{Conclusion}
\label{sec:conclusion}
In this research, we proposed a novel approach to financial portfolio optimization by leveraging the capabilities of quantum computing, specifically through the Quantum Walk Mixer integrated with a Quantum Approximate Optimization Algorithm (QWM-QAOA). We reformulated the traditional portfolio optimization problem into a knapsack problem, enabling us to address the NP-hard nature of this financial decision-making challenge. Our experimental evaluations, conducted across simulated, fake, and real quantum devices, demonstrated the effectiveness of our approach in yielding high approximation ratios, particularly in noise-free environments. The results of our study reveal that the QWM-QAOA approach is capable of producing efficient portfolio selections, with approximation ratios ranging from 100\% to 95\% in noise-free conditions and from 98\% to 70\% on noisy fake devices. However, when tested on real quantum hardware, the results showed a decline in performance due to noise and gate errors, reducing the approximation ratios to approximately 50\%. These findings highlight the current limitations of quantum hardware, particularly in dealing with noisy intermediate-scale quantum (NISQ) devices. While our proposed method showed promise, especially in noise-free environments, future work is needed to address the noise in real quantum devices. Error mitigation techniques with further exploration of quantum algorithms for optimization will be crucial in enhancing the robustness and scalability of this approach in real-world applications. Our study contributes to the growing field of quantum finance, offering insights into how quantum algorithms can be applied to complex financial problems such as portfolio optimization.

\section*{Code Availability}
The code that supports the findings of this study is openly available in the Github repository, \href{https://github.com/QCL-PKNU/QAOA-Knapsack-Portfolio-Optimization.git}{QWM-QAOA}.

\section*{Acknowledgment}
This research was supported by the "Regional Innovation Strategy (RIS)" through the National Research Foundation of Korea (NRF), funded by the Ministry of Education (MOE) (2023RIS-007).

\bibliographystyle{IEEEtran}
\bibliography{main.bib}

% \EOD

\end{document}